\documentstyle[aps,preprint,epsfig]{revtex}

\def \beq {\begin{equation}}
\def \eeq {\end{equation}}

\begin{document}

\draft

\title{Dynamical study of the singularities of gravity in the presence of 
non-minimally coupled scalar fields}

\author{L.R. Abramo$^1$,
L. Brenig$^2$,
E. Gunzig$^{2,3}$, 
Alberto Saa$^{2,4}$ \footnote{e-mails: 
{
abramo@fma.if.usp.br, 
lbrenig@ulb.ac.be, 
egunzig@ulb.ac.be, 
asaa@ime.unicamp.br}}}

\address{1)
Instituto de F\'\i sica, Universidade de S\~ao Paulo,
CP 66318, 05315-970 S\~ao Paulo, SP, Brazil
}

\address{2)
RggR, Universit\'e Libre de Bruxelles, 
CP 231, 1050 Bruxelles, Belgium.
}
\address{3)
Instituts Internationaux de Chimie et de Physique Solvay, 
CP 231, 1050 Bruxelles, Belgium. 
}

\address{4)
IMECC -- UNICAMP,
C.P. 6065, 13081-970 Campinas, SP, Brazil.}

\maketitle

\begin{abstract}
We investigate the dynamics of Einstein equations
in the vicinity of the two recently described 
types of singularity 
of anisotropic and homogeneous
cosmological models described by the action
$$
S=\int d^4x \sqrt{-g}\left\{F(\phi)R - \partial_a\phi\partial^a\phi
-2V(\phi) \right\},
$$
with general $F(\phi)$ and $V(\phi)$.
The dynamical nature of each singularity is elucidated, and we
show that both are, in general,
dynamically unavoidable, reinforcing the unstable
character of previous isotropic and homogeneous 
cosmological results obtained
for the conformal coupling case.

\end{abstract}

\section{Introduction}
We have 
recently\cite{PRD1} studied the singularities of 
homogeneous and anisotropic solutions of cosmological
models described by the action:
\beq
\label{act}
S=\int d^4x \sqrt{-g}\left\{F(\phi)R - \partial_a\phi\partial^a\phi
-2V(\phi) \right\},
\eeq
with general $F(\phi)$ and $V(\phi)$.
Such singularities appeared in the study of the robustness
of previously considered\cite{PRD} homogeneous and isotropic
solutions of cosmological models governed by (\ref{act})
with $F(\phi)=1-\frac{1}{6}\phi^2$, corresponding to 
the so-called conformal
coupling, and $V(\phi)=\frac{m}{2}\phi^2-\frac{\Omega}{4}\phi^4$.
These homogeneous and isotropic solutions present
some novel and interesting dynamical behaviors such as: 
superinflation regimes, 
a possible 
avoidance of big-bang and big-crunch 
singularities through classical birth of the universe 
from empty Minkowski space, spontaneous entry into and exit from inflation,
and a cosmological history suitable in principle 
for describing quintessence. The appearance of the
singularities implies that
these results are not robust, they are
radically changed, even for small disturbances
in initial conditions and in the model itself. 
We have shown that the singularities
are, essentially, of two types.
The first one corresponds to the hypersurfaces $F(\phi)=0$. 
It is not present in the
isotropic case, and it implies that all previous homogeneous
and isotropic solutions
passing from the $F(\phi)>0$ to the  $F(\phi)<0$ region are
extremely unstable against anisotropic perturbations.
The second type of singularity corresponds to $F_1(\phi)=0$, with
\beq
\label{f1}
F_1(\phi) = F(\phi)+\frac{3}{2}\left(F'(\phi)\right)^2,
\eeq
and it is present even for the homogeneous and isotropic cases. 
Although for small deviations
from the conformal coupling the latter singularities are 
typically very far from the region of
interest, in the general case they can alter qualitatively the global 
dynamics of the model due to restrictions that they impose on the
phase space. Again, the persistence of some of our previously described
results, in particular
the ones concerning
heteroclinic and homoclinic solutions, are challenged.

Both kinds of singularities have already been described 
before. To the best of our knowledge, Starobinski\cite{Starobinski} was the
first to identify the singularity corresponding to the hypersurfaces
$F(\phi)=0$, for the case of conformally coupled anisotropic solutions.
Futamase and co-workers\cite{Futamase} identified both
singularities
in the context of chaotic inflation in $F(\phi)=1-\xi\phi^2$ theories
(See also \cite{s2}).
The first singularity is always present for $\xi>0$ and the second one for 
$0<\xi<1/6$.
Our conclusions were, however, more general since we treated the case of
general $F(\phi)$ and our results were based on the analysis of
true geometrical invariants. Our main result is that the
system governed by (\ref{act}) is {\em generically} singular on both
hypersurfaces $F(\phi)=0$ and $F_1(\phi)=0$. Here, {\em generically}
means that it is possible to construct non-singular models if
one fine-tunes $F(\phi)$ and $V(\phi)$, as we have shown in \cite{PRD1}.
The physical relevance of such a fine-tuned model is still unclear.

As was shown in \cite{PRD1}, one can advance that there are some geometrically 
special regions in the phase space of the
model in question by a very 
simple analysis of the equations derived from the 
action (\ref{act}).
They are the Klein-Gordon equation
\beq
\label{kg}
\Box\phi - V'(\phi) +\frac{1}{2}F'(\phi)R=0,
\eeq
and the Einstein equations
\begin{eqnarray}
\label{ee}
F(\phi)G_{ab} &=& (1+F''(\phi))\partial_a\phi\partial_b\phi \nonumber \\ &-& 
\frac{1}{2}g_{ab}\left[ (1+2F''(\phi))\partial_c\phi\partial^c\phi 
+2V(\phi)\right] - F'(\phi)\left(g_{ab}\Box\phi - \nabla_a\phi\nabla_b\phi 
\right).
\end{eqnarray}
We consider now the simplest anisotropic homogeneous cosmological
model, the Bianchi type I, whose spatially flat metric is given by
\beq
\label{metric}
ds^2 = -dt^2 + a^2(t)dx^2 + b^2(t)dy^2 + c^2(t)dz^2. 
\eeq 
The dynamically relevant quantities here are
\beq
H_1 = \frac{\dot{a}}{a}, \quad H_2 = \frac{\dot{b}}{b}, \quad
{\rm and\ } H_3 = \frac{\dot{c}}{c}\ .
\eeq
For such a metric and a homogeneous scalar field $\phi=\phi(t)$, after
using the Klein-Gordon Eq. (\ref{kg}), 
Eq. (\ref{ee}) can be written as
\begin{eqnarray}
\label{ec1}
F(\phi)G_{00} &=& \frac{1}{2}\dot{\phi}^2 + V(\phi) - F'(\phi)\left( 
H_1+H_2+H_3\right)\dot{\phi}, \\
\label{e1}
\frac{1}{a^2}F(\phi)G_{11} &=& \frac{1+2F''(\phi)}{2}\dot{\phi}^2 -
V(\phi) - F'(\phi)\left( H_1\dot{\phi} + V'(\phi) -\frac{F'(\phi)}{2}R\right),
\\\label{e2}
\frac{1}{b^2}F(\phi)G_{22} &=& \frac{1+2F''(\phi)}{2}\dot{\phi}^2 -
V(\phi) - F'(\phi)\left( H_2\dot{\phi} + V'(\phi) - \frac{F'(\phi)}{2}R\right),
\\\label{e3}
\frac{1}{c^2}F(\phi)G_{33} &=& \frac{1+2F''(\phi)}{2}\dot{\phi}^2 -
V(\phi) - F'(\phi)\left( H_3\dot{\phi} + V'(\phi) - \frac{F'(\phi)}{2}R\right).
\end{eqnarray}
It is quite simple to show that Eqs. (\ref{e1})-(\ref{e3}) are
not compatible, in general,
 on the hypersurface $F(\phi)=0$.  Subtracting
(\ref{e2}) and (\ref{e3}) from (\ref{e1}) we have, on 
such hypersurface, respectively,
\beq
F'(\phi)(H_1-H_2)\dot{\phi} = 0,\ {\rm and\quad }
F'(\phi)(H_1-H_3)\dot{\phi} = 0.
\eeq
Hence, 
they cannot be fulfilled in general for anisotropic metrics. As it
was shown, this indeed corresponds to a geometrical 
singularity which cannot be prevented in general 
by requiring that $F'(\phi)=0$ 
or $\dot{\phi}=0$ on the hypersurface.

As to the second singularity we have, after taking the trace 
of the Einstein
equations, that:
\beq
\label{r}
R = R(\phi,\dot{\phi}) = \frac{1}{F_1(\phi)}\left(4V(\phi) +
3V'(\phi)F'(\phi) - (1+F''(\phi))\dot{\phi}^2 \right).
\eeq
Inserting Eq. (\ref{r}) in the Klein-Gordon Eq. (\ref{kg}), one
can see that it contains terms which are singular for $F_1(\phi)=0$. Again, 
as we will see, this corresponds
to an  unmovable geometrical 
singularity, and it cannot be eliminated, in general,
 by demanding that $F'(\phi)=0$ 
on the hypersurface $F_1(\phi)=0$. On both the  
hypersurfaces $F(\phi)=0$ and
$F_1(\phi)=0$ the Cauchy problem is ill-posed, since one cannot
choose general initial conditions.

The hypersurfaces $F(\phi)=0$ and
$F_1(\phi)=0$ also prevent the global definition of an Einstein frame
for the action (\ref{act}), defined by the transformations
\begin{eqnarray}
\label{eg1}
\tilde{g}_{ab} &=& F(\phi)g_{ab}, \\
\label{eg2}
\left(\frac{d\tilde{\phi}}{d\phi}\right)^2 &=& \frac{F_1(\phi)}{2F(\phi)^2}.
\end{eqnarray}
It is well known that in the Einstein frame the Cauchy problem is well posed. Again,
the impossibility of defining a global Einstein frame shed some doubts
about the general Cauchy problem.
Moreover, the standard perturbation theory for helicity-2 and helicity-0
excitations,
derived directly from Eqs. (\ref{eg1})-(\ref{eg2}), 
fails on both hypersurfaces\cite{G}.

The question to be addressed in the following sections
is the dynamical behavior of
the Eqs. (\ref{kg}) and (\ref{ec1})-(\ref{e3})  in the vicinity
of the two hypersurfaces corresponding to
$F(\phi)=0$ and $F_1(\phi)=0$. As we will
see, both hypersurfaces are dynamically unavoidable, meaning 
that they have an attractive neighborhood, excluding definitively the
possibility that these singularities are hidden by some dynamical
barrier that would prevent the solutions to reach them. Whenever a
solution enter in the attractive neighborhood, it will unavoidably
reach the singular hypersurface.

\section{Non-conservative systems and the divergence theorem}

For Hamiltonian systems, as a consequence of Liouville theorem,
phase space volumes are preserved under the system time
evolution. That means that if one chooses an initial closed 
hypersurface $S_0$ in the phase space and let each point
of $S_0$  evolve in time according to the system equations, the
closed hypersurface $S_0$ will evolve to another closed
hypersurface $S_t$ at some latter time $t$, and the volumes $V$
of the region enclosed by 
$S_0$ and $S_t$ are exactly the same, $V(0)=V(t)$. This is
a characteristic of conservative systems. The system given by
(\ref{kg}) and (\ref{ec1})-(\ref{e3}) is not conservative.
By choosing the set of coordinates $(\phi,\psi,p,q,r)$ (with 
$\psi = \dot{\phi}$,  $p=H_1+H_2+H_3$, $q=H_1-H_2$, and 
$r=H_1-H_3$),  for the phase space $\cal P$, Eq. (\ref{kg})
 and (\ref{ec1})-(\ref{e3})
 can be cast in the form
\beq
\label{E}
(\dot{\phi},\dot{\psi},\dot{p},\dot{q},\dot{r}) = \vec{W}(\phi,\psi,p,q,r).
\eeq
For the metric (\ref{metric}), we have the following
identities
\begin{eqnarray}
\label{gg}
G_{00} &=& H_1H_2 + H_2H_3 + H_1H_3, \nonumber \\
G_{11} &=& a^2\left( \dot{H}_1 + H_1(H_1+H_2+H_3) - \frac{1}{2}R\right),  
\nonumber\\
G_{22} &=& b^2\left( \dot{H}_2 + H_2(H_1+H_2+H_3) - \frac{1}{2}R\right), \\
G_{33} &=& c^2\left( \dot{H}_3 + H_3(H_1+H_2+H_3) - \frac{1}{2}R\right),  
\nonumber\\
R &=& 2\left( \dot{H}_1 + \dot{H}_2  + \dot{H}_3  
+ {H}_1^2 + {H}_2^2 + {H}_3^2 + 
 H_1H_2 + H_2H_3 + H_1H_3
\right).  \nonumber
\end{eqnarray}
Using them, the components of $\vec{W}$ can be
explicitly computed from Eqs. (\ref{ec1})-(\ref{e3}),
\begin{eqnarray}
\label{W}
W_\phi &=& \psi  \nonumber \\
W_\psi &=& -p\psi - V'(\phi) +\frac{1}{2}F'(\phi)R(\phi,\psi),  \nonumber\\
W_p &=&  -\left[ (F(\phi)+2F'(\phi)^2)p^2+
\frac{3}{2}(1+2F''(\phi))\dot{\phi}^2 - 3V(\phi) - 3 F'(\phi)V'(\phi) 
\right. \nonumber \\
 & & \left. - p\dot{\phi}F'(\phi) 
  + (F(\phi)+F'(\phi)^2)(q^2 + r^2 - qr) \right] /(2F_1(\phi)) 
\nonumber\\
W_q &=&  - \left(p+\frac{F'(\phi)}{F(\phi)}{\psi}\right)q, \nonumber\\
W_r &=&   - \left(p+\frac{F'(\phi)}{F(\phi)}{\psi}\right)r. 
\end{eqnarray}
The divergence theorem assures us that the volume $V$ of a closed
hypersurface $S_t$ of $\cal P$ evolves in time as:
\beq
\label{T}
\dot{V}(t) = \int_{S_t} ({\rm div}\, \vec{W})\, d\rm vol,
\eeq
where the integral is performed in the region enclosed by $S_t$.
The divergence of the vector field $\vec{W}$ determines, therefore,
how fast a volume of a closed hypersurface of $\cal P$ is
expanded $ ({\rm div}\, \vec{W}>0)$ or contracted  $({\rm div}\, \vec{W}<0)$. 
For conservative systems, one has  ${\rm div}\, \vec{W}=0$.
A straightforward calculation here give us
\begin{eqnarray}\label{div}
 {\rm div}\, \vec{W} &=& -p -2\left(p+\frac{F'(\phi)}{F(\phi)}{\psi}\right)
\nonumber \\
&-& \frac{\left(F(\phi)+2F'(\phi)^2\right)p  
+ \left( F'(\phi) \left( 1+F''(\phi) \right)  - 
\frac{1}{2}F'(\phi) \right)\psi }{F_1(\phi)}
\end{eqnarray}
It is clear from (\ref{div}) that our system suffers violent
contractions and/or expansions in the neighborhood of the
hypersurfaces $F(\phi)=0$ and $F_1(\phi)=0$. Let us consider
each of them separately since they lead to different kinds
of singularity.

\section{Phase space contraction and expansion near $F(\phi)=0$}

Homogeneous and isotropic solutions, for whose $q=r=0$,
are known to be perfectly regular on $F(\phi)=0$\cite{PRD}, in contrast
to (\ref{div}) that presents unequivocally a divergence on
this hypersurface. A closer analysis of the vector field
$\vec{W}$ (\ref{W}) reveals  that the divergent contractions and
expansions near $F(\phi)=0$ are associated to the directions
$q$ and $r$, namely the quantities that measure the anisotropy
of the solution. The other directions of the flux defined by
$W$ are regular on the hypersurface $F(\phi)$. That means
that if a solution is perpendicular to the divergent directions
$q$ and $r$, solutions for which $q=r=0$, {\i.e.}
isotropic solutions, it will evolve without suffering any
violent contraction or expansion in the other directions
of $\cal P$. Since $W_q$ and $W_r$ are proportional to
$q$ and $r$, respectively, an initially isotropic solution $q(0)=r(0)=0$
remains isotropic for all latter $t$, $q(t)=r(t)=0$.
We can say that isotropic solutions are orthogonal to the
divergent fluxes.  However, any amount, no matters
how small, of anisotropy (non vanishing $q$ or $r$) will break
the orthogonality and the solution will prove the divergent
directions, being strong contracted or expanded. This is the
dynamical origin of the instabilities of anisotropic solutions
near $F(\phi)=0$.

Let us suppose now that $F'(\phi)\ne 0$ on the hypersurface
$F(\phi)=0$ (if $F'(\phi)$ vanishes on the hypersurface $F'(\phi)=0$, 
then by Eq. (\ref{f1})
both hypersurfaces $F(\phi)=0$ and $F_1(\phi)=0$ coincide). The corresponding
pole on $\phi_0$ $(F(\phi_0)=0)$
in the volume integral (\ref{T}) will have as numerator the
factor $-2F'(\phi_0)\psi$, implying that the flux defined by (\ref{W})
passes from a catastrophic contraction to a catastrophic expansion
as one passes by $\phi_0$. Since $W_\phi=\psi$ (see Eq. (\ref{W})), any
solution approaching the hypersurface $F(\phi)=0$ with $\psi\ne 0$
(we excluded from the analysis the possibility of having
fixed points on $F(\phi)=0$, for these cases, of course, it makes
no sense to talk about ``crossing''  $F(\phi)=0$) will cross it and,
hence, prove the divergent phases of contraction and expansion.

In the expanding ``side'' of the hypersurface $F(\phi)=0$,
$q$ and $r$ diverges as $\phi\rightarrow\phi_0$, and the system
will be unavoidably driven toward a spacetime singularity\cite{PRD1},
as we can conclude by considering, for instance, 
the Kretschman invariant $I=R_{abcd}R^{abcd}$, which
for the metric (\ref{metric}) is given by
\beq\label{kret}
I = 4\left(
\left(\dot{H}_1+H_1^2\right)^2 + 
\left(\dot{H}_2+H_2^2\right)^2 + 
\left(\dot{H}_3+H_3^2\right)^2 
+ 
H_1^2H_2^2 + H_1^2H_3^2 + H_2^2H_3^2
\right).
\eeq
The invariant $I$ is the sum of non negative terms.
Moreover, any divergence of the variables $H_1$, $H_2$, $H_3$, or of
their time derivatives, would suppose a divergence in $I$,
characterizing a real geometrical singularity. Since 
the relation between the variables $p$, $q$, $r$, and
$H_1$, $H_2$, $H_3$ is linear, any divergence in $p,q,r$ 
or of their time derivative, will suppose a divergence
in $I$.

\section{Phase space contraction and expansion near $F_1(\phi)=0$}
 
The hypersurface $F_1(\phi)=0$, as the previously one considered in
the last section, also separates regions of catastrophic contraction
and expansion in the phase space $\cal P$. However, in contrast 
to the previous one, this hypersurface leads to singularities even
for homogeneous and isotropic solutions. The divergent directions
now are $p$ and $\psi$, and there are no solutions of (\ref{E})
orthogonal to them, since, in contrast to the $q$ and $r$ directions,
for which $W_q$ and $W_r$ are respectively proportional do $q$ and $r$,
$W_\psi$ and $W_p$ do not vanish for $\psi=0$ and $p=0$. In this case,
no solution can escape from crossing $F_1(\phi)=0$. In the expanding
side of $F_1(\phi)=0$, $p\rightarrow\infty$ as $\phi\rightarrow\phi_1$
($F_1(\phi_1)=0$), implying the divergence
of the invariant (\ref{kret}).

\section{Final remarks}

The singularities described in the precedent section imply that
the model presented in \cite{PRD,IJTP1} is not robust, since 
its main conclusions were a 
consequence of very especial initial conditions, {\i.e.} they
are valid only for solutions orthogonal (namely the isotropic
ones) to the divergently
expanding directions $q$ and $r$.
For instance, all homogeneous and isotropic solutions crossing the
$F(\phi)=0$ hypersurface are extremely unstable against anisotropic
perturbations. Any deviation 
from perfect  isotropy
(expressed by nonvanishing $q$ and $r$ variables) for these solutions, 
however small, will lead catastrophically to a
geometrical singularity. Many of the novel dynamical behaviors
presented in \cite{PRD,IJTP1} depend on these solutions. This is 
the case, for instance, of some solutions exhibiting 
superinflation regimes. The heteroclinic and homoclinic
solutions identified in \cite{PRD,IJTP1} can cross the 
$F(\phi)=0$ hypersurface and, hence,
they also suffer the same
instability against anisotropic perturbations. The homoclinic
solutions were considered as candidates to describe a non-singular 
cosmological history, with the big-bang singularity being
avoided through a classical birth of the universe from empty
Minkowski space. Apart from $F(\phi)=0$ singularities, 
these solutions are also affected by the singularities of the type 
$F_1(\phi)=0$.
Suppose that the conformal coupling is disturbed by a very
small negative term: $F(\phi)=1-(\frac{1}{6}-\epsilon)\phi^2$.
The $F_1(\phi)=0$ singularities will be near the 
$\phi=\pm 1/\sqrt{\epsilon}$ hypersurfaces. Although they are
located far from the $F(\phi)=0$ regions, they alter the
global structure of the phase-space. In this case, they restrict
the existence of homoclinics, rendering a non-singular 
cosmological history more improbable.

The singularities do not affect the conclusions obtained
by considering solutions inside the $F(\phi)>0$ region.
The asymptotic solutions presented in \cite{IJTP1}, for
instance, are still valid. The conclusion that
for large $t$ the dynamics of any solution (inside $F(\phi)>0$)
tends to an infinite diluted matter dominated universe 
remains valid. Moreover, for small anisotropic deviations
($q$ and $r$ small in comparison with $p$), 
the
solutions inside $F(\phi)>0$, 
for large $t$, approach exponentially isotropic matter-dominated universe.

A singularity-free model can be constructed by demanding a well
behaved $\rm div\,\vec{W}$ on both hypersurfaces. This can be
achieved\cite{PRD1} by requiring
$F(\phi_0) =F'(\phi_0) = 0$, and by choosing a $V(\phi)$ that goes to 0 
at a proper rate when $\phi\rightarrow\phi_0$. Moreover  
$F_1(\phi)$ must have no other zeros than the ones of  $F(\phi)$.
Models for which $F(\phi) = \zeta\phi^{2n}$ and
$V(\phi) = \alpha\phi^{2(2n-1)} + {\rm\ high\ order\ terms}$, 
for instance, fulfill these
requirements. However,
such a highly fine-tuned class of model is of no physical interest here, 
since it does not contain $F(\phi)>0$ and $F(\phi)<0$ regions and consequently
has no solution for which the effective gravitational constant
$G_{\rm eff}$ changes it sign along the
cosmological history. The stability of such solutions were the
starting point of the analyses of the pioneering work \cite{Starobinski}
and of the present one as well.

\acknowledgements

The authors
wish to thank G. Esposito-Farese for previous discussions concerning the
Cauchy problem for models with $F(\phi)=0$, and
Profs.  Annie and Albert Sanfeld for the warm hospitality
in Mallemort, France, where this work was initiated.
They also acknowledge the financial support from the EEC 
(project HPHA-CT-2000-00015), from OLAM - Fondation pour la Recherche
Fondamentale (Belgium), 
from Fondation Science et Environnement (France),
and from CNPq (Brazil).


\begin{references}
\bibitem{PRD1} L.R. Abramo, L. Brenig, E. Gunzig, and A. Saa,
Phys. Rev. {\bf D67}, 027301 (2003).

\bibitem{PRD} E. Gunzig, A. Saa, L. Brenig, V. Faraoni, T.M. Rocha Filho, and 
A. Figueiredo, Phys. Rev. {\bf D63}, 067301 (2001);

\bibitem{IJTP1}A. Saa,  E. Gunzig, 
 L. Brenig, V. Faraoni, T.M. Rocha Filho, and 
A. Figueiredo, Int. J. Theor. Phys. {\bf 40}, 2295 (2001). 

\bibitem{Starobinski} A. A. Starobinsky, P. Astron. Zh. {\bf 7}, 67 (1981)
[Sov. Astron. Lett. {\bf 7}, 36 (1981)]. 

\bibitem{Futamase} T. Futamase and K. Maeda, Phys. Rev. {\bf D39}, 
399 (1989); T. Futamase, T. Rothman, and R. Matzner,
 Phys. Rev. {\bf D39}, 405 (1981).


\bibitem{s2} S. Deser, Phys. Lett. {\bf 134B}, 419 (1984);
Y. Hosotani, Phys. Rev. {\bf D32}, 1949 (1985);
O. Bertolami, Phys. Lett. {\bf 186B}, 161 (1987).

\bibitem{G}G. Esposito-Farese and D. Polarski, Phys. Rev. {\bf D63},
063504 (2001).



\end{references}
\end{document}